# Integration of RFID Applications in a Web B2B Platform for Enterprise Supply Networks

Turcu, Cristina
Prodan, Remus
Cerlincă Tudor
Cerlincă Marius
Turcu, Cornel
Popa, Valentin
Goloca, Alexandru

1st October 2007

"Stefan cel Mare" University of Suceava
13, University Street, RO-720229 SUCEAVA
cristina@eed.usv.ro

**Abstract**

B2B applications focus on using the Internet and/or extranet to improve business-to-business partnerships and transform inter-organizational relationships. RFID is relatively low-cost data and wireless transmission technology that helps manufacturers to improve a number of business applications and processes. In this paper we present an RFID_B2B system that brings together the B2B and RFID advantages and which could be a viable solution for the potential problems created due to the globalization process. Using the developed system may help customers sharpen data accuracy, process supply chain transactions faster, and improve supply chain and inventory management.

Keywords:  *RFID, B2B, mobile application, tag, control gates.*

## 1. Introduction

### 1.1. B2B Concept

Business to business (B2B) exchange generally refers to any business transaction occurring between two separate business entities. This includes the exchange of both products and service. The term is most commonly used in connection with e-commerce and advertising, when it is targeting businesses as opposed to consumers. Examples of exchanged products and services might include the selling of raw material inputs from one firm to another, the sale of capital equipment, the purchasing of commercial insurance or the contracting of one firm with another for the procurement of accounting services.

Some of B2B systems benefits are:
- a low total cost of ownership; this is a result of the easy configuration of always-changing, complex partner networks and relationships;
- a high and secure visibility into partnering activities and performance; this can be reached at all points in the network;
- an increased partner mind-share; this can be obtained through revenue-generating tools providing; this tools can be used independently by a partner or in collaboration with other partners or the vendor;
- a lower cost for marketing and selling;
- a shorter selling cycle;
- just in time delivery; this is one of the most important advantages of B2B and enables the company to have the track of good with the help of electronic commerce.

Automated business-to-business transactions are not an entirely new concept. Large organizations have been using automated systems for a number of years, and some have been programmed to exchange business transactions with other automated systems as far back as the early nineties [1]. But, user testing shows that B2B websites have substantially lower usability than mainstream consumer sites. According to Jakob Nielsen [2], the major problems with B2B sites are:
- the fail in supporting customers' decision-making process by preventing them from getting the information they need to research solutions;
- they use segmentation that don't match the way customers think of themselves;
- they lack pricing information (the users in the study prioritized prices as the most critical type of information).

Our research team developed an RFID_B2B system that consists in a viable and efficient solution to eliminate these problems.

## 1.2. RFID Technology

Radio frequency identification (RFID) is a relatively new automatic identification and data capture (AIDC) technology that uses digital data encoded into a radio tag (or "smart label") that is collected by a reader using radio waves. RFID is similar to another AIDC technology, bar code technology, but instead of optically scanning bar coded labels it uses radio waves to capture data from tags and no direct line of sight is required for this data exchanged between the tags and the readers. The key components of any RFID system, tags, are made up of three parts:

- microchip: holds the desired information, e.g., information about the physical object to which the tag is attached;
- antenna: transmits information to a reader using radio waves;
- packaging: encases the previous components (chip and antenna) to permit attaching the tag to the desired physical object.

Tags use a variety of power sources:
- *active tags* - contain their own power source (a battery), that is used to run the microchip's circuitry and to broadcast a signal to a reader;
- *passive tags*, that have no internal power source. Instead, they draw power from the reader;
- *semi-passive tags*, that use a battery to run the chip's circuitry, but communicate by drawing power from the reader.

Passive tags are undoubtedly less expensive than active tags and most companies are focusing on passive tags.

RFID technology is emerging as a powerful and proven tool for streamlining production at manufacturing facilities of all sizes [3].

## 2. RFID@B2B

Our research team implements an RFID_B2B system that brings together the B2B and RFID advantages and which in the near future could be a viable solution for the potential problems created due to the globalization process. Thus, the RFID_B2B system refers to the business relations in large enterprises, corporations and groups, as regards the control of the materials along their entire supply chain. The system suggests applying the RFID technology by using RFID 13.56 MHz High Frequency (HF) passive tags to identify materials and assemblies. Thus, based on the ID codes of the materials and assemblies, it is possible to control the content and the origin of any finite product, the content of assemblies and the origin of any constituent component, and so on, for each company which contributed to the creation of the finite product. By extending the system to the entire supply-chain - final producer, supplier, the manufacturer's suppliers, etc. - the customer can follow the course of materials included in the final product, up to the primary sources. In order to accomplish this, all the necessary tracking information will be comprised in the tags attached to the materials, assemblies and finite products.

### 2.1. General presentation

The presented system is a very complex one. The research team chooses to design a layered architecture arranged in such a way that the lower layers support and enable the upper layers. This architecture has some advantages: divide the complex system into several more manageable components, allow different groups to work on different layers concurrently etc. The RFID_B2B system is structured on three levels: the corporation level, the local level and data collection level at the material control departments (Figure 1).

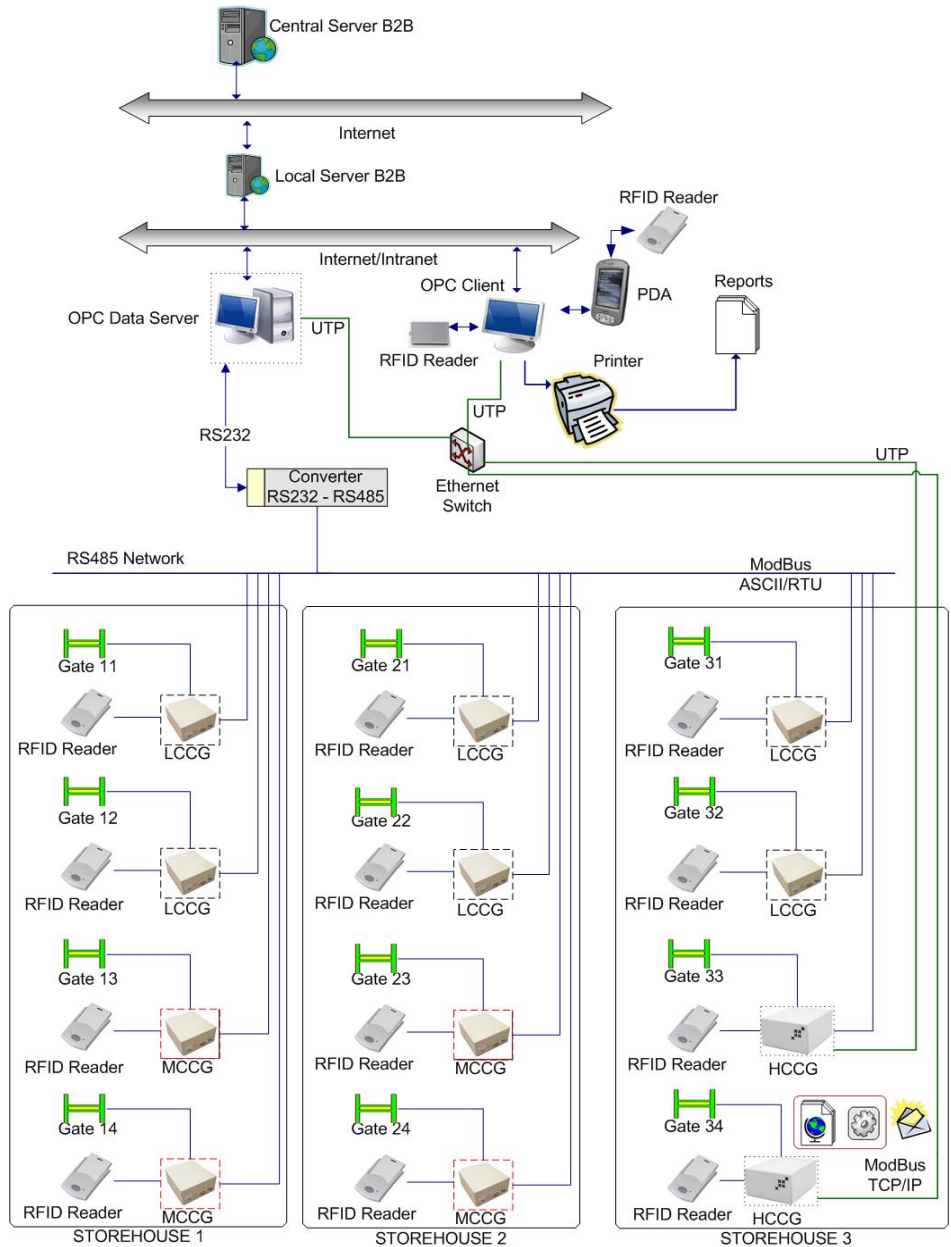

Figure 1. The system architecture

At the platform's *corporation level,* the following achievements have been made:
- services are offered to ensure the support for concluding contracts, along with the agreements, the conventions at the board level of the corporations, the firm groups or the representatives of the firm groups;
- informational management of the group/corporation enterprises, ensuring supervision of the material flows (with effects on the establishing of supply and sales strategies), as well as of the good functioning of the internal network of the group/ corporation;
- supply of reports regarding the current activities within the group/ corporation, including alarm in case of generation of specified events.

The following are provided at the *local level* or at the enterprise level:
- proper administration of the received, sent, defective, repaired, returned entities at the enterprise level;
- access to the company servers network, as well as communication management along the supply- sales main chain, providing the opportunity to manage and access the information referring to the route followed by materials, assemblies and finite products;
- coordination of the materials/ assemblies flow in order to ensure adequate distribution to corresponding departments, as well as to deliver the order to the gates in departments;
- documents delivery for controlling the production, materials, finite products, assemblies, including those in the service department.

Different applications of RFID are implemented in the *data collection level* in order to write and read the data from the tags attached to the materials, assemblies and finite products. At this level, the communication is wired or wireless.

As for the development environments and the SGBD employed to create the components and the application, Microsoft Visual Studio 2005 was considered to be the most appropriate solution. As for the databases, we chose Sybase SQL Anywhere 10 for the PDA devices and Microsoft SQL Server 2005 for the PC database server.

## 2.2. PC applications

All PC applications present a high degree of generality that permits a simple implementation in various activity fields without any modifications in the structural level of software applications. Thus, the user can define a template that describes the data format to be used for writing data into tags. Through an advanced template editor (Figure 2) the user can establish necessary fields (e.g. acquisition date, location, current value) and their type (character, string, integer, real). When a client orders a particular item from a supplier, that supplier sends an order confirmation and template information back to the client. The client company stores that information in its system locally.

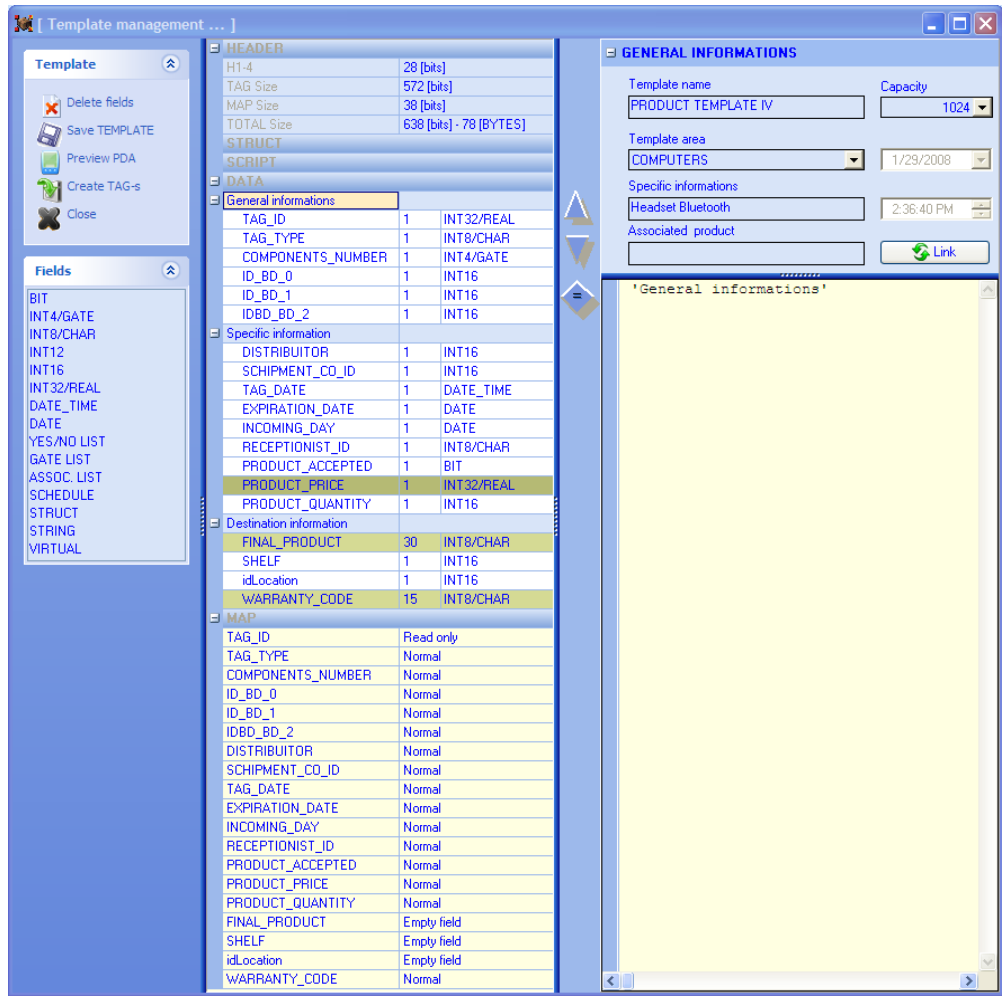

Figure 2. The tag template editor

All templates are created at the PC level and then transferred to the PDA through specialized web services. An important aspect is related to the visual organization of the fields on a tag so that they can be read on the PDA display. The visual space on the PDA touch screen is far too small and it is rather difficult to create/update a tag that has too many fields; the low display resolution and small display screen inhibit information to be displayed completely and clearly. That is why, users can define at the PC their own visual areas according to their needs and then group all tag fields. In general, each group will consist of several fields with the same purpose. Figure 3 exemplify the visual organization of the fields on a tag. All visual areas created at the PC level are then transferred to the PDA (Figure 4).

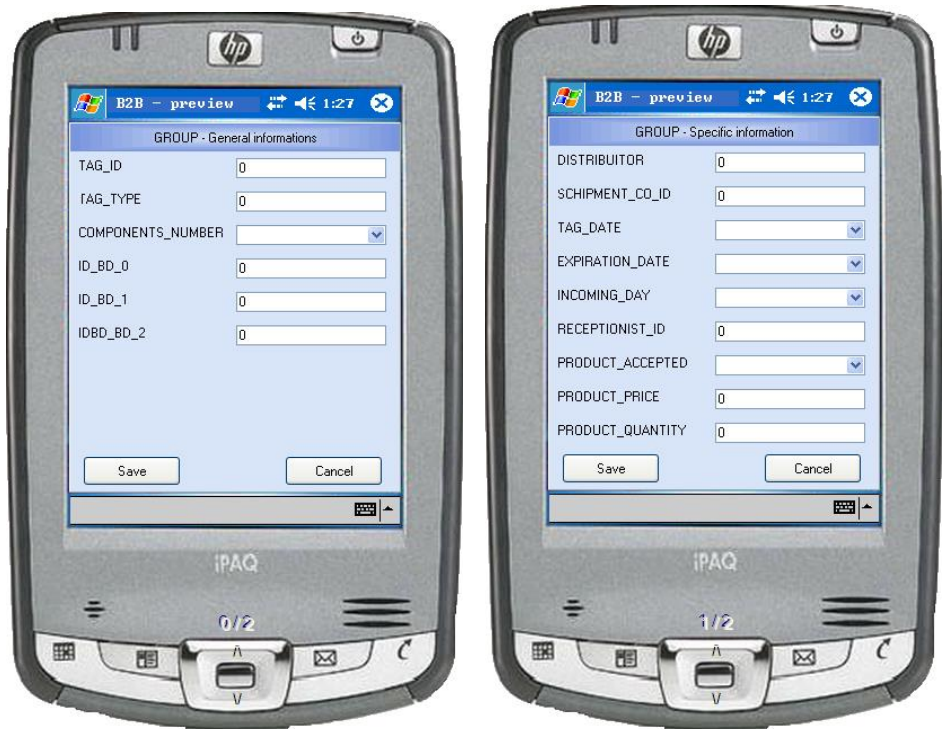

Figure 3. The preview of visual organization of tag fields

The PC applications from the data collection level enable RFID data capture, filtering, correlation and database update. Also, the PC-control gates communication is enabled for sending the commands to the control gates, respectively for PC database updating with control gates read information.
The PC applications installed at local and corporation level provide a service-oriented architecture (SOA) platform consisting on distributed application server, secure reliable messaging, local database, systems management. Thus, the RFID_B2B system can be the service provider by enabling some functions as web services, which can be accessed by external systems. Also, the RFID_B2B is a service requester by enabling it to invoke web services hosted by external systems (e.g. currency converter).

### 2.3. Mobile applications

The implemented RFID_B2B system also handles multiple PDA devices and PC

servers and facilitates data sharing among these devices. Thus, mobile application performs the following functions:
- read and write RFID tags;
- enable the management of the database, which stores information related to tags;
- work in stand-alone mode (independently of the main servers);
- store huge data;
- integrate and exchange information with complex RFID_B2B systems and other PDA mobile devices;
- enable the management of system registered users;
- employ a multi-user and user-friendly interface (Figure 4).

Figure 4. PDA visual organization of tag fields

### 2.4. RFID-based control gates

Our system permits the use of three types of RFID-based control gates, which vary both in price and the package of features that are being offered. The three types are: Low Complexity Control Gates (LCCGs), described in [4], Medium Complexity Control Gates (MCCGs), described in [5], and High Complexity Control Gates (HCCGs), described in [6]. All three types of control gates have the ability:
- to read fields from RFID tags and save them into a non-volatile history memory using a file system;

- to send stored data to a PC (only at request);
- to modify fields on RFID tags as a result of a command from a PC or as the result of an internal script execution;
- to fire alarms as the consequence of some actions and events that occurred;
- to control relays and read data using digital inputs and outputs.

All three kinds of control gates are able to use a RS485 serial connection to communicate with a PC through ModBus communication protocol and a RS232 connection to communicate with the RFID reader.

### 2.5. System benefits

The presented system offers a high degree of flexibility and helps companies of all sizes enable their customers to do business on demand — when they want, where they want and how they want. Other system benefits are:
- Assures realtime inventories so the users can always receive accurate, up-to-date inventory information;
- Offers the possibilities to share meaningful data with supply chain partners;
- Permits strengthening customer and partner relationships with collaboration;
- Speeds and simplifies the deployment and management of e-commerce sites;
- Maximizes performance, scalability and adaptability of partners systems;
- Provides rich, ready capabilities for products catalog and content management;
- Permits a greater visibility through realtime product updates, availability and pricing information;
- Offers personalization capabilities.

## 3. Further developments

The following aspects might be taken into account as future directions for development:
- integration of intelligent agent technology, through the development of some intelligent agents, which allow the defining of the user's profile, the collecting of information and its filtering (considering the criteria chosen by users), etc;
- application development for mobile wireless equipments (m-commerce);
- on-line processing of transactions (banking-financial-accounting) involved in B2B exchanges.

## 4. Conclusions

RFID is relatively low-cost data and wireless transmission technology that can help companies to improve the business to business processes. The presented system helps small, medium and enterprise organizations to improve productivity and provide better service to their customers by providing a flexible solution for all of a company's B2B needs. Given slim profit margins, companies are looking for ways to save on costs while remaining globally competitive. RFID@B2B may be their answer.

## 5. Acknowledgment

This work was supported in part by the Romanian Ministry of Education and Research under Grant named "Integration of RFID Innovative Applications in a Web B2B Platform for Enterprise Supply Networks - RASMEN" 06CEEX I 03/2005.